# The Application of Autocorrelation SETI Search Techniques in a ATA Survey

Short title: SETI Autocorrelation Survey with ATA

G. R. Harp[1], R. F. Ackermann[1], Alfredo Astorga[1], Jack Arbunich[1], Jose Barrios[1], Kristin Hightower[1], Seth Meitzner[1], W. C. Barott[2], Michael C. Nolan[3], D. G. Messerschmitt[4], Douglas A. Vakoch[1], Seth Shostak[1], J. C. Tarter[1]

[1] Center for SETI Research, SETI Institute, 189 Bernardo Ave., Ste. 100, Mountain View, CA, 94043.

[2] Electrical & System Engineering Dept., Embry-Riddle Aeronautical University, 600 S. Clyde Morris Blvd, Daytona Beach, FL, USA, 32114.

[3] Arecibo Observatory, HC3 Box 53995, Arecibo, PR 00612, USA.

[4] Department of Electrical Engineering and Computer Sciences, University of California at Berkeley, 387 Soda Hall, Berkeley, CA 94720-1776.

**Abstract**

We report a novel radio autocorrelation (AC) search for extraterrestrial intelligence (SETI). For selected frequencies across the terrestrial microwave window (1-10 GHz) observations were conducted at the Allen Telescope Array to identify artificial non-sinusoidal periodic signals with radio bandwidths greater than 4 Hz, which are capable of carrying substantial messages with symbol-rates from 4-$10^6$ Hz. Out of 243 observations, about half (101) were directed toward sources with known continuum flux > ~1 Jy over the sampled bandwidth (quasars, pulsars, supernova remnants, and masers), based on the hypothesis that they might harbor heretofore undiscovered natural or artificial, repetitive, phase or frequency modulation. The rest of the targets were mostly toward exoplanet stars with no previously discovered continuum flux. No signals attributable to extraterrestrial technology were found in this study. We conclude that the maximum probability that future observations like the ones described here will reveal repetitively modulated emissions is less than 1% for continuum sources and exoplanets, alike. The paper concludes by describing a new approach to expanding this survey to many more targets and much greater sensitivity using archived data from interferometers all over the world.

# 1 Introduction

Searches for Extraterrestrial Intelligence (SETI) at radio frequencies traditionally focus on slowly modulated narrowband signals (Cocconi & Morrison 1959; Drake 1961; Oliver & Billingham 1971; Shuch 2011; Tarter 2001). The premise of the narrowband (~1 Hz) search is that relatively weak narrowband ETI signals may be present but hidden in ordinary astronomical observations.

An unspoken assumption is that all strong (> 1 Jy) radio sources already known to astronomers have a natural origin. This paper recognizes that this statement is not fully supported by existing observations. Some well-known strong radio sources might harbor a hidden message masquerading as, or piggybacking on, a strong natural source. While many pulsars and other sources may have been previously tested for repetitive power modulation, the authors are not aware of previous work testing for encodings that use e.g. constant-power phase modulation from known bright sources. The latter is the focus here.

Suppose ET were to construct a powerful transmitter sending information at a bit rate between $10^3 - 10^9$ Hz. To most radio telescopes, such a transmitter is indistinguishable from a natural continuum source because the time fluctuations are too short to appear in standard detectors. However, these same signals can be detectable by autocorrelating the electric field amplitude and phase, otherwise known as a field autocorrelation (FAC) detection. Here we present what we believe to be the first radio search for ETI using FAC detection of complex signals. As an alternative we also consider intensity autocorrelation for detection of incoherent amplitude-modulated signals.

Recently there has been a resurgence of theoretical research on searching for *wideband* engineered signals that may be used for interstellar messaging (Gardner & Spooner 1992; Harp et al. 2010a, 2015; Von Korff et al. 2013; Messerschmitt 2012; Messerschmitt & Morrison 2012; Morrison 2012, 2017; Siemion et al. 2010). Meanwhile, techniques developed for very long baseline interferometry have been adapted to capture substantial bandwidths (>1 MHz) of digitized time series data for SETI post-processing (Harp et al. 2010a, 2015; Korpela et al. 2001; Morrison 2012; Siemion et al. 2010; Tarter et al. 2010; Wayth et al. 2011). The benefit of archiving such data is that they may be processed in ways that could not be performed in real time. This paper reports SETI observations that make use of this non-traditional approach.

## 1.1 Conventional Matched Filter Bank Searches and AC

Searches for narrowband continuous and pulse signals depend upon the assumption of preconceived signal types. The prototypal ET signal appears as a narrow sloping or slightly curved trace in a frequency vs. time plot or waterfall plot (c.f. Figure 1). The intensity-inverted waterfalls of Figure 1 portray a narrowband signal from the ISSE3 spacecraft (left) and a dispersed pulse of radiation from the Crab pulsar (right). To highlight the similarity of the waterfalls, the space and frequency axes are swapped between left and right images. The ISEE3 signal is not vertically aligned because of the relative acceleration between spacecraft and detector. The pulse is similarly slanted because light propagating in the cold plasma of the interstellar medium is dispersed (increasingly retarded at lower frequencies).

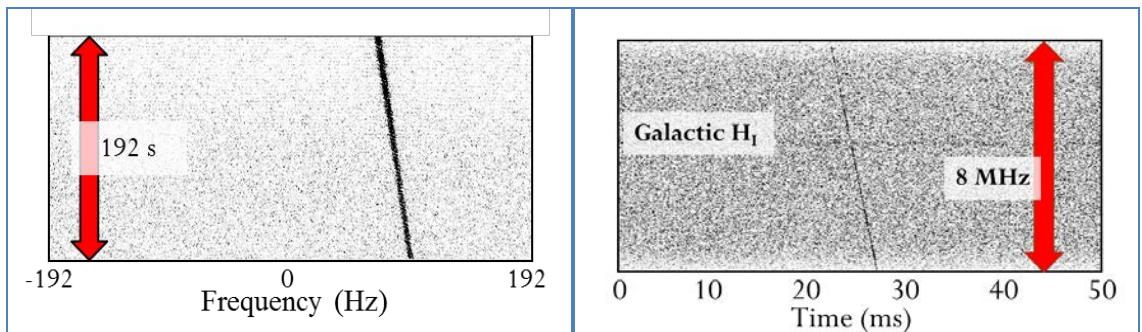

**Figure 1: (Left) Time versus frequency waterfalls taken at ATA where the narrow dark trace indicates detection of a narrowband transmission from spacecraft ISEE3 during its closest approach to Earth. (Right) Frequency versus time waterfall where the trace indicates detection of one wideband "giant pulse" from the Crab pulsar. The horizontal line in the crab data corresponds to galactic HI emission at 1420 MHz.**

Searches for such signals look for modestly curved traces in the time-frequency domain having small perpendicular cross-section (width << cross-width). Such searches can be parameterized with 2 to 3 independent parameters: the frequency $f$, slope or drift rate $df/dt$, and sometimes curvature or $d^2f/dt^2$. In an equivalent pulse search, the time and frequency variables are swapped: $t$, $dt/df$, $d^2t/df^2$.[1]

Usually, signal searches use a matched filter bank (MFB) which effectively prepares a set of test waveforms spanning the parameter ranges and compares them to the observed waveform. The benefit of MFBs is their sensitivity. Historically it has been argued that it is better to search a narrow parameter space with the greatest sensitivity than to search a wider parameter space with moderate sensitivity (Oliver & Billingham 1971).

In this campaign, we use AC detection rather than an MFB. The search philosophy is that AC searches for wide bandwidth[2] signals and can piggyback on MFB narrowband searches requiring only nominal extra processing. For weak signals, AC detection is less sensitive than narrowband detection for the same ET transmission power. Yet AC is sensitive to a different and larger class of signals that generally do not appear in narrowband searches. Thus we broaden the possibilities for detection of an artificial signal with only a small additional computational expense. We argue that in future searches, it makes sense to implement both narrowband and AC detectors for a more effective search, especially when observing with radio interferometers.

## 1.2 Research Hypotheses

To summarize, this observational campaign tests the following hypotheses (with more detail to be found in the discussion section):

**Hypothesis 1.** The emitted electric fields of many previously discovered strong radio sources with flux >1 Jy are modulated with a repeating pattern either directly by extraterrestrials or due to some heretofore unknown physics.

**Hypothesis 2.** Many exoplanets emit moderate bandwidth (e.g. 1 MHz) artificial signals that were not previously discoverable in continuum surveys or in narrow band ETI searches. If such signals contain repetitive structure, they can be detected by autocorrelation.

To the best of our knowledge, these hypotheses cannot be excluded based on past observatations, and the results here represent a new foray into the search for extraterrestrial intelligence.

This survey focuses only on moderate bandwidth repetitive signals with waveforms substantially more complex that pure sinusoids. The SETI Institute already has a sensitive ultra-high resolution spectrometer used for narrowband searches but instrumental limitations prevented the pursuit of a commensal, narrowband search along with the AC search. Low-resolution power spectra were generated from all the captured data to verify data quality, but the frequency resolution of these power spectra was not good enough to pursue a narrowband search.

---

[1] In practice, a fast search for naturally dispersed pulses can take advantage of the fact that the second and third parameters are co-dependent thus requiring only a 2 parameter matched filter bank in the search (Zackay, B. & Ofek, E. O. 2012).

[22] What is considered wide-bandwidth is relative to the very narrow bandwidth signals of conventional searches. In this paper, AC signals with bandwidth up to 7 MHz as compared with typical 1 Hz bandwidth signals in conventional SETI.

Besides the source types mentioned above, this campaign includes a small number of targets with special interest for SETI such as the galactic anti-center and the Earth-Sun Lagrange L4 point, and many reference observations used as comparators for testing the direction of origin of newly detected candidate signals.

The rest of this paper contains a description of the autocorrelation signal detection techniques used here and compares their sensitivities to traditional matched filtering. Within this discussion we introduce the details of the observations, followed by a few examples of man-made technological signals that appeared in the search. An analysis section summarizes the results and extracts quantitative results. This is followed by a brief demonstration of how repetitive signals can be detected in archived radio interferometer data that might have been overlooked in previous analyses. We demonstrate this idea with new observations of intentional, repetitive signals reflected from the moon, and how this research can be extended to the analysis of large libraries of archived radio interferometer data.

## 2 Signal Detection Algorithms

Autocorrelation-based detection strategies are a useful addition to the toolkit of the traditional SETI narrowband search. Autocorrelation 1) is sensitive to a wide class of signals that are not effectively detected in a narrowband search, 2) has the potential for detecting signals containing messages with substantial information rates[3], and 3) is insensitive to dispersion in the interstellar medium (Harp et al. 2010a) since identical signals subject to the same dispersion remain identical. The last point is crucial since wideband radio signals are strongly dispersed in the interstellar medium and this can otherwise obfuscate the detection of wideband communication signals.

We demonstrate this immunity to dispersion with a simulation. We model a distant transmitter emitting four short pulses of sinusoidal radiation at 1 GHz and emitted over a period of 4000 seconds as in Figure 2 left (solid line). After traveling 1600 light years through a plasma with mean electron density equal to that of the galactic interstellar medium (0.01 e$^-$ cm$^{-3}$), the dispersion measure (DM) of the signals will be about 5 pc cm$^{-3}$. Applying the well-known formula for cold plasma dispersion, an estimate of the dispersion measure is (Cordes 2002) delay (s) $= 0.00415 \, DM/f^2 \, \text{(GHz)}$. We calculate the pulse (electric field) amplitudes as they would appear at the receiver, Figure 2, left (green dots). After dispersion, the pulses are broadened and have lower peak intensity. (During the time period near 2000 seconds two of the dispersed pulses overlap and coherently interfere, but this has no impact on subsequent pulse reconstruction.) On the right-hand side of Figure 2, we display the complex-valued field autocorrelation (FAC, algorithm defined below) of the received signal with itself (lavender line). As expected, the FAC detector shows 6 spikes for the original 4 pulses (zero delay spike has been suppressed). FAC response is compared to another method, intensity autocorrelation (green dots, IAC, defined below).

---

[3] **For example, information encoded using a finite alphabet of symbols with repeating elements at more or less random delays.**

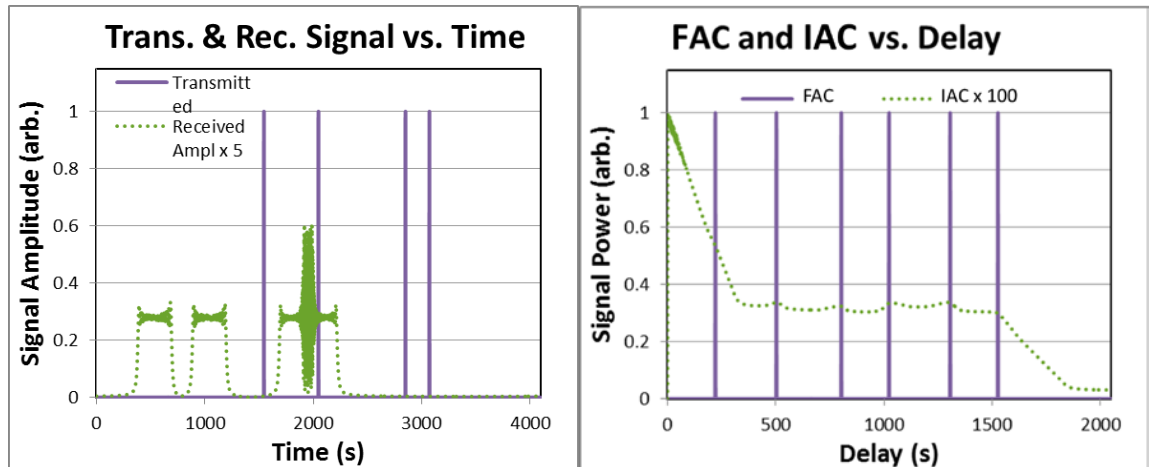

**Figure 2: Transmission and reception of a coherent pulsed signal after passage through the interstellar medium for 1600 years. The left graph compares a baseband copy of the transmitted electric field amplitude at the transmitter (lavender solid line) and receiver (green dots). On the right we plot the autocorrelation of the received electric field (FAC, line) and autocorrelation of intensity (IAC, dots) for the signal. The repetitive signal is detectable in both FAC and IAC.**

The example of Figure 2 visualizes our general point: An FAC filter detects a coherent[4] repetitive signal passed through any stationary linear filter just the same as if the data are not filtered. Hence FAC is immune not only to dispersion but also resistant to the effects of slowly varying scintillation (timescales longer than observation period) in the interstellar medium.

FAC discriminates between natural and engineered signals. For the engineered coherent pulses modeled in Figure 2, FAC discrimination (height of peaks compared to surrounding values) is generally quite good provided the received signal has a sufficient signal to noise ratio (see below). By comparison, pulsars and quasars emit light that is not coherent over long periods of time and which would are not detected with FAC. This is why FAC is not used for pulsar searches. As far as we know, only artificial, engineered processes can give rise to signals detectable in FAC.

The coherent FAC process used here is not ordinarily accessible for astronomy at optical frequencies, which usually measures intensities. Intensity autocorrelation can be thought of as the extension of FAC methods to light curves (i.e. flux versus time). Intensity autocorrelation finds application in many areas of physics, such as characterization of ultra-short optical pulses, the detection of weak optical pulsars (Leeb et al. 2015), interstellar scintillation of quasars (Rickett et al. 2002), and measuring stellar rotation periods from Kepler light curve data (McQuillan et al. 2014).

---

[4] In this discussion, "coherent" pulses are defined as having strong correlation between separate pulses That is, pulses in the train have identical electric field waveforms.

Incoherent signals can be detected by intensity autocorrelation (IAC) even when FAC fails to find them. IAC is closely related to the method of synchronous averaging of power used in pulsar searches. While the basic principles behind IAC are not new in radio astronomy, we briefly describe it as a comparison to FAC.

## 2.1 Signal Types

In SETI searches at the ATA we identify a few archetypal quasi-stationary signal waveforms:

(A) *Narrowband*: A nearly-sinusoidal signal where the coherence time is of order 100 ms or longer. In the workhorse implementation called SonATA (SETI on ATA), such signals may carry of order 10 bits/s of information. We refer to this method as "conventional SETI."

(B) *Wideband power detection*: With an interferometer like the ATA, it is straightforward to capture "snapshots" of the sky, covering many square degrees. These snapshots can be compared across time periods from seconds to hours, or even longer. Such power intensity images may be examined for point sources that could be associated with ET transmitters or other unexpected radiation.

(C) *Cyclostationary*: A finite alphabet of radio waveforms representing symbols is transmitted in a time series. FAC, then, is sensitive to any pair of like symbols, repeated at specific delays. This mode is often used in satellite communication implementing error correction (Gardner & Spooner 1992; Leshem et al. 2000; Morrison 2012, 2011). Message information might be encoded in the transmission center frequency, time delay or signal phase. When substantial galactic dispersion is present, such symbols will generally overlap in time (c.f. dispersed pulses in Figure 2) which significantly impacts signal discovery in IAC but has little impact on signal detectability with FAC.

(D) With phase modulation, an alphabet may be constructed from a single symbol multiplied by an overall phase factor. With such alphabets, all symbols correlate with all other symbols (see e.g. GPS example in (Harp et al. 2010a)). However, destructive interference between correlations of different symbol pairs can dilute FAC sensitivity. One solution to this problem is described in (Morrison 2012) where, for each delay in the plot, a separate computation is performed for each symbol period and the correlation magnitudes are summed incoherently (rather than as complex numbers in ordinary FAC). We do not adopt that method here because it requires several orders of magnitude greater computation time and was not feasible here.

As the number of symbols in the alphabet grows, FAC sensitivity is also diluted since the average density of like symbols is reduced. This effect does not substantially alter the conclusions of this paper.

On Earth, cyclostationary signals which contain no extrinsic information (same symbol repeated over and over) are sometimes used in radar applications (e.g. Arecibo planetary radar) to generate a controlled wide bandwidth signal. If such a signal were discovered in SETI, it would still be conspicuously artificial and of interest.

(E) *Amplitude Modulation (AM)*:
  a) *Direct transmission,* as with broadcast radio. AM is more prone to errors caused by noise and fading than frequency or phase modulation.
  b) *Receive, delay and transmit (RDT)*: One suggestion is that ET may take advantage of a strong natural source to enhance detectability. For example, ET can set up a transmitter on a line of sight between Earth and a quasar. ET collects the quasar signal, amplifies it, and re-transmits a delayed copy toward Earth (Harp et al. 2010a). Information may be embedded in such a signal using a time-dependent delay.
  c) *Source modulation*: Direct amplitude modulation of an astronomical source might be accomplished by modulating or pumping the source itself. As proof of the principle, (Weisberg et al. 2005) has identified a maser source that is modulated by a pulsar. Since the light crossing time of the smallest masers is on the order of hundreds of seconds (Boyd & Werner 1972) modulations at shorter timescales indicate a localized pumping source.
  d) *Other:* The signal modulations above assume that the received signal is approximately stationary over time periods on the order of 100 ms or smaller. Of course, an infinity of other signal types exist including non-stationary signals that might be picked up with our autocorrelation detectors below. This is not a problem since any signal we find has a high likelihood of being artificial.

## 2.2 Sensitivity of Various Detectors

We consider a single polarization system where the telescope voltages are digitally sampled and a total of $N_s$ samples are gathered in the observation. The total power $P_t$ for a 1σ detection of a signal is proportional to the telescope's system equivalent flux density (SEFD), that is, when the astronomical flux density is equal to the system noise power. More precisely,

$$P_t = \frac{(\text{SEFD})}{\sqrt{N_s}}, \qquad \text{where} \qquad \text{SEFD} = \frac{2\eta k_B}{\eta_{eff} A} T_{sys}, \qquad (1)$$

and $\eta < 1$ is the power loss factor associated with digitization, $k_B$ is Boltzmann's constant, $A$ is the total collecting area and $\eta_{eff} < 1$ is aperture efficiency, or the fraction of $A$ that is captured by the receiver in the combined optical system. At the ATA, $\eta_{eff} = 0.6$ (Harp et al. 2011). For equal signal in two polarizations, $P_t$ is divided by $\sqrt{2}$.

Using measured antenna system temperatures, Table 1 displays the computed sensitivity measures for the various detection algorithms. At four frequencies used in observations, we compute the SNR for a 10-minute observation of a 1 Jy source (e.g. ETI transmitter or a quasar) emitting uniformly over the 7 MHz detector bandwidth in column 3. The 4th column shows the flux of a source detectable at 1σ for a total-power measurement of a wideband signal. For a narrowband signal, column 5 shows the minimum detectable flux in a 1 Hz bin. Using FAC and IAC detectors and under the favorable assumptions of Gaussian noise (Fano 1951) and that the sample period is commensurate with the symbol period, column 6 applies.

Table 1:

Detection Thresholds for Various Algorithms
Assumes 600 s at 7 MHz bandwidth, $A_{eff} = 438$ m².

| $f$ (GHz) | Tsys (K) | SNR 1 Jy Source | 1σ Wideband Detectable Flux (Jy) | 1σ Narrowband Detectable Flux in 1 Hz bin (Jy) | 1σ FAC, Detectable Flux (Jy) |
|---|---|---|---|---|---|
| 1.43 | 80 | 129 | 0.009 | 27 | 0.037 |
| 3.04 | 120 | 86 | 0.014 | 40 | 0.054 |
| 6.667 | 95 | 108 | 0.011 | 32 | 0.043 |
| 8.4 | 137 | 75 | 0.016 | 45 | 0.063 |
| $f$ | Tsys | SNR | Ratio of Total power at receiver for equivalent detectability | | |
| 1.43 | 80 | 129 | 2,000 | 1 | 10,000 |

The 4th column in Table 1 shows the power detection threshold for a signal whose bandwidth $BW_{signal} \geq BW_{detector}$. The analogous detection threshold for a very narrowband signal is larger by a factor of $\sqrt{N_{chan}}$, where $N_{chan}$ is the number of channels across the detector bandwidth. This is because all the power in the threshold wideband signal must be transmitted in a single channel. The detection thresholds for a 1 Hz bandwidth signal are shown in column 5 of Table 1 for reference only. When comparing sensitivities of different detectors, column 5 is the relevant column for an arbitrary MFB irrespective of the actual transmitted signal bandwidth. The sixth column shows the detection sensitivity of the FAC and IAC. The most relevant comparison for column 6 is column 4. We return to this comparison in the discussion section.

In Table 1 bottom row we multiply the threshold sensitivities by their effective bandwidth, which allows us to compare total transmitter power for equivalent cases. The ratio of transmitter powers are 1:2000:10000 for the narrowband, wideband and FAC, respectively to generate the same detected signal with each metric. As expected, the narrowband signal, where the recipient knows the signal shape, requires the least total power. The wideband power is larger because of the fact that for each antenna pair we are correlating two independent measures of the signal, both of which contain uncorrelated noise. The FAC power is larger still since both the reference and test signals must both be present at the same time in a single measurement, leaving less power for either of them.

This paper focuses on FAC as an alternative to the narrow band search for ETI. In the next section we describe FAC predict its relative sensitivities.

## 2.3  The Power Spectrum and Two Alternative Detectors

The power spectrum (PS), FAC and IAC detectors are introduced with notional expressions for their computation as compared with that for the traditional power spectrum (PS) in Eqs. (2), (3), and (4), respectively.

The power spectrum (PS) uses an ordinary fast Fourier transform and this statistic represents a typical MFB detector[5] where the matching functions are oscillating exponentials. The FAC detector is sensitive to repeating modulations of the electric field. The IAC detector is sensitive to repeating modulations of the received power. Using FAC, amplitude modulation or constant amplitude signals with phase modulation or frequency modulation are detectable. IAC is sensitive to power modulation but not constant-amplitude (e.g. phase) modulation. Comparatively, typical CCD detectors for light measure only the received power, after which IAC processing is sometimes possible. Radio measurements are special in that the phase of the electric field can be measured, permitting FAC detection.

The actual digital calculations require subtle corrections that are described in the Appendix. For the given observation time the number $N_s$ of digitized samples are broken into $N_b$ equal blocks of length $M$. Each block is processed with a Fourier transform (FT), which is then averaged over all blocks:

$$PS(f) = \sum_{n=0}^{N_b-1} \left| \sum_{m=0}^{M-1} \exp(-i\,2\pi f\, t_m)\, s(nT + t_m) \right|^2 \tag{2}$$

$$FAC(\tau) = \sigma_s^{-2} \left| \sum_{l=0}^{M-1} \exp(i\,2\pi f_l\, \tau) \sum_{n=0}^{N_b-1} \left| \sum_{m=0}^{M-1} \exp(-i\,2\pi f_l\, t_m)\, s(nT + t_m) \right| \right| \tag{3}$$

$$IAC(\tau) = \sigma_{s^2}^{-2} \sum_{l=0}^{M-1} \exp(i\,2\pi f_l\, \tau) \sum_{n=0}^{N_b-1} \left| \sum_{m=0}^{M-1} \exp(-i\,2\pi f_l\, t_m) \left| s(nT + t_m) \right|^2 \right| \tag{4}$$

---

[5] In actual observations, signals that deviate slightly from pure sinusoids are also checked.

In Eqs. (2)-(4), $s(nT + t_m)$ is the value of the sample at point m in the $n^{th}$ block, $t_m$ is the time associated with $s$ measured from the start of the $n^{th}$ block, $f_i$ is the baseband frequency, $\tau$ is the time delay measured from the start of the $n^{th}$ block and $\sigma_s^2$ and $\sigma_{s^2}^2$ are the respective mean square variances of the field and intensity. These equations define the power spectrum (PS), averaged magnitude of sampled field autocorrelation (FAC), and averaged magnitude of sampled intensity autocorrelation (IAC) detectors, in that order (note squaring operation inside FT for IAC). As shown, the AC algorithms are implemented using the Wiener-Khinchin convolution theorem (Wozencraft & Jacobs, 1990 p. 73) with forward and inverse Fourier transforms. The units of PS and FAC are power, while IAC is different. With IAC, the fluctuations of the measured power are the "signal" we wish to characterize. The units of IAC are power also, based on this definition of "signal."

In application we apply a simple 5σ threshold to the FAC statistic to identify signals that are likely to be engineered. Detection is done by visual inspection of graphs. As a function of frequency, approximate 1$\sigma$ threshold flux values for detection are given in Table 1.

## 2.4 Observational Design

The ATA is a dual-polarization 42-element interferometer located in Northern California, comprising 6.1 m dishes, dual-linear polarization feeds can that operate in 4 simultaneous frequency bands centered anywhere between 1-10 GHz (Welch et al. 2009).

The signals from many ATA antennas are delayed and summed in a beam former (Barott et al. 2011). Complex-valued (8-bit real, 8-bit imaginary, hence $\eta \approx 1$) samples from the beam former are collected at a rate of 8.73 MS/s. An anti-aliasing bandpass filter before the digitizers limits the effective bandwidth to 7 MHz. The phased array beam diameter can be estimated as $0.1°/f$, where $f$ is the observation frequency in GHz. Quasars are effectively point-sources for the ATA and many are more than 100 times more powerful than the background level in the time period over which we perform FAC.

Data collections were made over approximately 14 months (Jan. 2010 – Mar. 2011). The typical ATA configuration used 25 of the 6.1 m antennas in a phased array beam on the source of interest. We accumulated data for typically 600 s, for an effective total of $N_s = 4.2 \times 10^9$ samples per observation. Referring to Equation (3), $M = 2^{23}$ or approximately 1 second long. The sampled data from each observation were reduced using Gnu Octave (Eaton et al. 1997) to compute the various statistics.[6] All the source data are freely available in an internet archive (Harp et al. 2010b). Plots of PS and IAC were also generated for reference, but not included in the analysis here since those detectors have been well characterized before.

---

[6] Source codes available upon request.

Many source targets (69) were chosen because they are known to have fluxes > 1 Jy. From Table 1 we see that if the entirety of that flux were repetitively modulated, then it would appear with high signal to noise in this survey. Additionally, sources for examination were chosen as follows: Exoplanets and Kepler objects of interest (59). Observations of 13 pulsars (flux > 1 Jy) and the galactic anti-center were performed based on the hypothesis that ET might broadcast in the direction opposed to a known source in their field of view as an aid for our detection of them. Two strong methanol masers (>>1 Jy at the spectral line) were observed to see if those masers might be artificially modulated. Six O type stars were chosen to study the hypothesis that advanced extraterrestrials might set up beacons using bright stars as an energy source. Other special pointings include the sun and moon for artifact tests, the Earth-Sun Lagrange L4 point (where transmitters might be left in a stable orbit), and the ecliptic North Pole.

Observational frequency bands were selected to have minimal strong radio frequency interference (RFI) and often encompass certain "magic" frequencies such as the $H_I$ line (1.420 GHz), $\sqrt{2}\, H_I$ (2.008 GHz), $2\, H_I$ (2.840 GHz), $\pi\, H_I$ (4.462 GHz), the methanol maser line (6.667 GHz) and 8.4 GHz because it is close to the upper limit of ATA's receivers.

Only positive delays were examined since the curves are symmetric about zero delay. Inspection was performed over the delay range $1 \times 10^{-4} < \tau < 0.25$ s, where the bias due to finite $M$ is limited to no more than a factor of 2. The lower limit is chosen to exclude features due to delays in time of arrival at different antennas in the ATA. The time spacing between plotted delays is the same as for the original sampling $\Delta \tau = 1.4 \times 10^{-7}$ s.

Interesting signals were noted and if those signals appeared in observations of more than one spatial direction, they were identified as a ground or space-based man-made source.

## 2.5  Identification and Elimination of Human-Generated Signals

It is the goal of this paper to focus attention on detecting technological signals no matter how they are generated. We wish to avoid the tedious work of identifying signals' specific transmitters, be they radar, cell phone, satellite, etc. Instead, we use direction of arrival methods to classify signals as either coming from the vicinity of Earth or coming from farther away. Using these methods we can reliably eliminate signals from human transmitters out to about 10% of the orbit of the moon without any prior knowledge of spectrum usage. This is important since there is no part of the radio spectrum free from human interference. Since public documentation regarding most transmitters is spotty at best, it is impossible to rule out most signals out by any means other than direction of arrival.

# 3  Results

We demonstrate the utility of FAC detection with a few examples of (interfering) signals detected in this survey in Figure 3. Figure 3A shows a fiducial observation taken in the direction of a known spacecraft beyond lunar orbit (Deep Impact) in the frequency range of its communication downlink. This 35 Jy signal[7] is exactly the type of signal we hope to observe. Plots B and C show an example of an unidentified 7 Jy signal that appeared in multiple pointings, hence identified as interference. Plot D shows how transmission from strong satellites (in this case Geostationary Operational Environmental Satellites or GOES) generate interfering (but not imaging) signals at 3 Jy[8] even when the telescope is not pointed at them.

Plot E shows another unidentified 2 Jy signal which we labeled as interference because we deem it unlikely that an extraterrestrial transmitter would coincidentally repeat at such a round number (10.000 ms) in Earth units (also seen in more than one direction). Plot F shows a weak unidentified signal near the left-hand side with power at our $5\sigma$ detection limit (0.32 Jy for this frequency). This signal was observed in multiple pointing directions.

Taking a closer look, the signal in Plots B and C occurs right in the so-called protected radio astronomy band for the $H_I$ line. This signal was not visible in IAC, indicating that it is probably phase-modulated. The FAC signal was observed in 7 MHz bands from 1400-1470 MHz with approximately uniform power. From this it is easy to understand why there was no detectable feature in the power spectrum (PS) for this signal. Also, because the source of this signal does not resolve to a point on the sky, even a repetitive signal as strong as this is not easily detected in ordinary interferometer observations and went undetected at the ATA for nearly a decade.

---

[7] Effective flux if spread over full 7 MHz bandwidth of our observations, computed with Table 1: and cross checked against an earlier DSN measurement
https://descanso.jpl.nasa.gov/DPSummary/di_article_cmp20050922.pdf.

[8] A GOES satellite uses a 7 W transmitter, an antenna with 11 dB gain, and typical downlink bandwidth of 100,000 Hz. Therefore, in its frequency of emission it has a flux density >1,000,000 Jy, so the residual signal we see is well within our expectations.

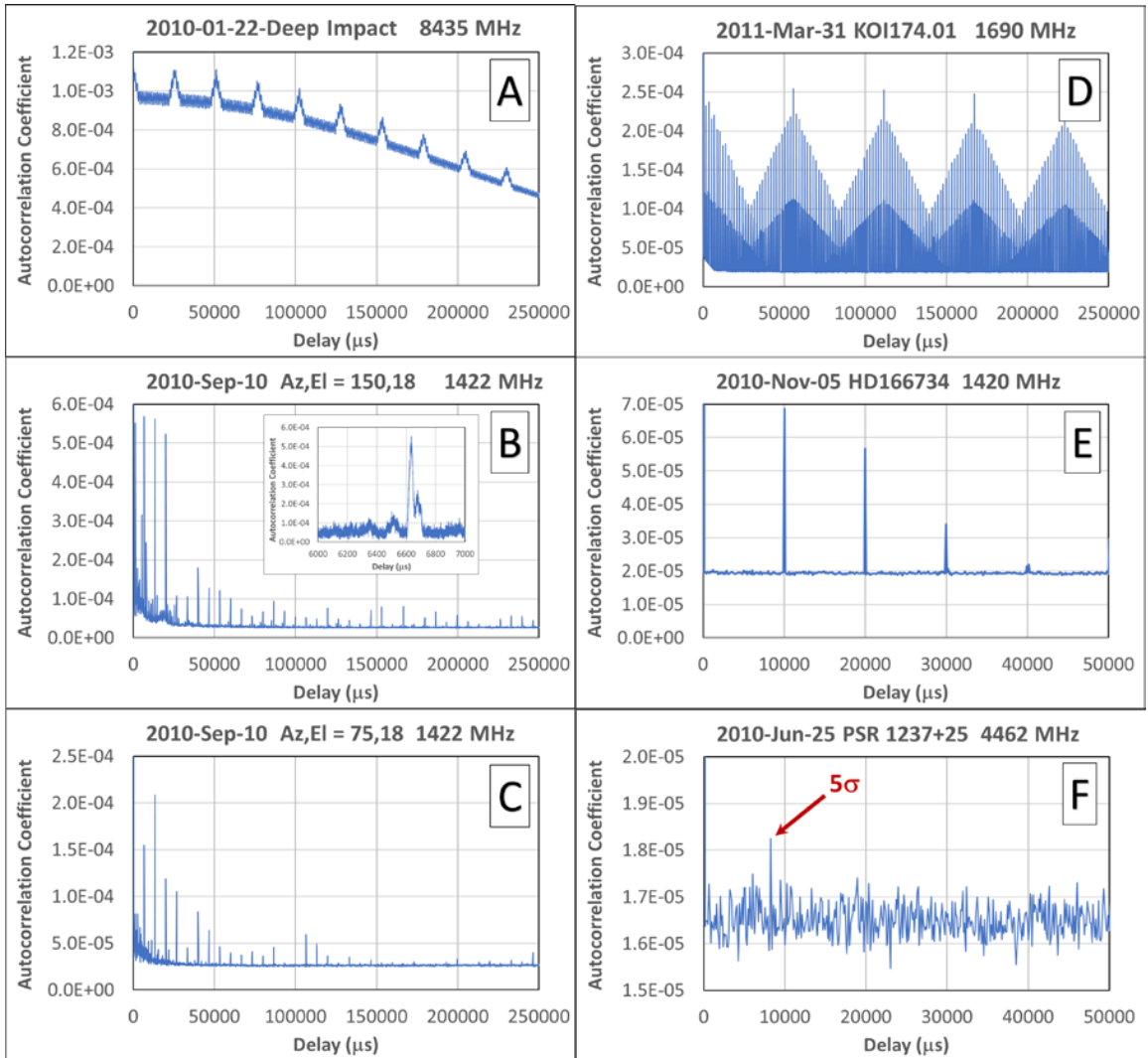

**Figure 3:** FAC plots of power as a function of delay on various sources. A: The Deep Impact spacecraft in downlink transmission band. B and C: An unexpected interfering signal observed in multiple pointing directions. The inset in B shows a blow up of the delay region around 6500 µs. D: While pointed at a Kepler object of interest, an observation at 1690 MHz shows signal in the frequency range of Geostationary Operational Environmental Satellites, even though all GOES satellites were far from the pointing direction. E: Observed while pointed at a blue supergiant star, the signal period of 10.000 ms is identified as human-generated due to its close alignment with Earth-based time units. F: While pointed at a pulsar, one of the smallest signals detected in this paper. Although this too is interference, this plot gives an impression of the signal to noise ratio obtained in these measurements.

We estimated the coherence time of the plots B and C signal by computing the statistic using different sample block lengths $M$, but keeping all of the data. The first FAC peak had a maximal plateau for $M$ corresponding to block times greater than 140 ms, while smaller $M$ resulted in a smaller peak. This plateauing behavior indicates that the repetition period is drifting slowly during the observation, changing by the delay of a single sample (1.2 $\mu s$) in about 140 ms. This demonstrates that we can detect repetitive signals even if they are weakly non-stationary.

The origin of this signal has not been identified nor do we speculate on its origin. This study shows how interfering signals are distinguished from true ETI transmissions using direction-of-arrival estimation. This is important; while some interfering sources can be identified with known transmitters, it is more often the case that the frequencies of such human-generated signals do not correspond with any cataloged source of radiation. Interfering signals are ubiquitous below 5 GHz, and once identified as interference there is no need for further investigation.

The interesting signal in plot D shows how satellite communications can generate interference in FAC observations. We intentionally observed in the transmission band of the GOES weather satellites while pointing far away from any of those satellites. Such strong satellite transmissions make radio observations impossible at frequencies where they occur.

Plot F is provided to highlight the background noise level of all observations. Because the correlations are computed over a finite period, we expect an average autocorrelation coefficient of ~$10^{-5}$ even when the input data contain only noise.

Summarizing all the observational results: We found 10 distinct signals in FAC that had no counterparts in PS. At least ten times as many unique signals were observed in the power spectra which indicates that today, interfering signals are more likely to be narrowband than repetitive. All unidentified signals were observed multiple times and in different pointing directions, indicating they arise from terrestrial/satellite communications. No unidentified FAC signal was observed with center frequency above 5 GHz. This indicates very little RFI over the entire upper half of the terrestrial microwave window (~1-10 GHz), frequencies where new future SETI observations will be least impacted.

These examples demonstrate a couple of points: 1) FAC is demonstrated to detect repetitive signals that will not be found in a conventional SETI search, and 2) FAC is a good detector for local RFI that may be difficult to characterize otherwise. We do not know of another radio observatory that uses FAC as a means of identifying RFI and we suggest that this approach may be useful elsewhere.

# 4 Analysis

Now we consider the research question[9], "What is the highest probability for the existence of true repetitive ETI signal that is compatible with our results?" We emphasize that since this is the first substantial survey looking for constant-power repetitive signals, we have little *a priori* knowledge about this question.

### 4.1.1 Hypothesis 1

Because there is little previous work on FAC surveys, it is difficult to quantify the state of our knowledge at the beginning of this survey. For rhetorical purposes, we introduce two hypotheses that encapsulate our goals.

---
[9] A more expansive description of our research hypotheses is given in the Introduction.

We hypothesize that most previously discovered strong radio sources with flux >1 Jy are constantly modulated with a repeating pattern either directly by extraterrestrials or due to some heretofore unknown physics.

The total population of possible independent observations of all sufficiently strong sources can be estimated as follows. We refer to the NRAO VLA Sky Survey (Condon et al. 1998), which used the VLA to image and identify sources of continuum radiation from most of the sky (except near south pole). The NVSS reports more than 2000 sources with flux greater than 1 Jy. Multiplying by the solid angle of the sky dived by the complete solid angle observed in the survey (and including strong masers, etc.), we estimate the existence of about 2690 sources with flux $\geq 1$ Jy. The frequency search window, (1-10 GHz) can be divided into 1286 independent 7 MHz ranges as are used here. The number of similar observations required to survey every 1 Jy source from 1-10 GHz is about $4 \times 10^6$.

Combining observations of bright quasars, pulsars and masers into a single group, we have 89 independent observations on 23 sources > 1 Jy. We assume a uniform distribution of repetitively modulated sources in position and frequency. We call the probability that a signal will be detected in any one trial $p_{hit}$, and the converse probability that no signal is detected $p_{miss} = 1 - p_{hit}$. What value of $p_{Hit}$ is consistent with our observations at a 99% confidence level? We write

$$P(89 \text{ misses}) = (1 - p_{hit})^{89} > 100\% - 99\% = 1\% . \qquad (5)$$

Solving for hit probability gives $p_{hit} < 5\%$, or no more than 1 in 20 sources are being modulated. If we wished to decrease $p_{hit}$ to be less than one in a million, we would require about four million negative observations. We set this as an aspirational goal.

It is worth noting that our entire survey severely under-samples the full source search space in the frequency range of 1-10 GHz where the source modulation may be present (sampling 89 frequency samples out of 3.5 million). Therefore our probability estimates here and below should be greeted with some skepticism. The reason we include numerical estimates at all is that we wish to establish limiting values that can be used for comparison in future surveys.

We reiterate this result including all caveats: Of the estimated ~2690 radio sources in the sky with flux > 1 Jy in the radio frequency range 1-10 GHz, no more than 5% of them might bear (either natural or intentional) repetitive modulation with repeat periods between $1 \times 10^{-4} < \tau < 0.25$ s.

### 4.1.2 *Hypothesis 2*

We suppose that most stars harboring exoplanets are sending repetitive signals. As we now know that most stars harbor exoplanets, this distinction is not as important as we once thought.

With 88 distinct observations, the math is almost identical to that for hypothesis 1. We conclude that at a 99% confidence level, no more than 5% of such stars are actively transmitting repetitive signals that are detectable in this survey.

Table 2:

New Limits on Probability of Repetitive ET Transmitter

| Source Type | Num Distinct sources and distinct frequencies | Num Possible Observations 1-10 GHz | $p_{Hit}$ |
|---|---|---|---|
| Distinct Source > 1 Jy | 89 | $4 \times 10^6$ | 0.05 |
| Exoplanet Stars | 88 | $\approx \infty$ | 0.05 |

Finally, another result worthy of comment: The L4 Lagrange point showed no evidence for an artifact transmitting detectable signals in 3 observations at 1420, 2008 and 3991 MHz.

## 4.2 Number of Signal Waveforms Probed by Narrowband vs FAC

Given $M$ samples, it is always possible to represent an arbitrary waveform with a basis of $M$ test waveforms. But when searching for signals we wish to choose a basis set that concentrates all the signal power into a small number of the basis functions so they stand out above the noise. To illustrate this point, consider that a very narrow pulse spreads equal power across every bin in a PS. In the presence of noise, PS is therefore a poor pulse detector. But a pulse basis, where each basis function is merely a linear combination of sinusoids, will readily detect pulsed signals. Different detectors are more or less sensitive to different linear combinations of the same set of basis signals and, as we shall see, the number of *distinguishable* waveforms tested by one method can be larger than the number of independent waveforms.

Given $M$ time series samples, it is interesting to compare the number of distinguishable waveforms that are tested by PS and FAC. Neglecting possible drifts, a fast Fourier transform is a matched filter bank for the set of all sinusoidal signals with periods between 2 and M time intervals. Hence a power spectrum analysis tests an observed time series for $N_W(\text{PS}) = (M-1)$ potential signal waveforms.

Surprisingly, a FAC analysis probes a larger number of waveforms, despite being derived from the same number of samples. This is because for each FAC period $P_n$ characterized as being $n$ samples in duration, there are a number $(n-1)$ of linearly independent waveforms that are detected. To avoid over counting, we establish a lower limit on the number of distinct waveforms $N_W(\text{FAC})$ by choosing only periods where $n$ is a prime number. Thus,

$$N_W(\text{FAC}) > (2+3+5+7+\ldots+M-1) \approx \int_2^{M-1} dx \frac{x}{\ln(x)} > M^{7/4} > N_W(PS), \quad (6)$$

where the integral takes advantage of the prime number theorem. The power law $M^{7/4}$ is an empirical fit that works well for $M < 2^{23}$, the number used here. Hence for $M$ samples, FAC tests for more waveforms than PS. It is important to be clear in our meaning of the "number of tested waveforms" in this context. Naturally, there really are only $M$ (truly) independent waveforms obtainable from $M$ samples. $N_W$ is a measure of the number of distinct waveforms for which FAC concentrates the signal power into just one or a few bins as an aid for signal detection. This is of value only when compute resources are limited. The price paid for added flexibility is that FAC requires a larger signal to noise ratio than PS to obtain the same fixed threshold of detector sensitivity (see last row in Table 1).

But the real motivation behind FAC detection is that we wish to be sensitive to at least some information-carrying signals, which by definition are not sinusoidal waveforms. Classic and recent theoretical studies of the interstellar medium as a communication channel (Drake 1965; Messerschmitt 2012; Messerschmitt & Morrison 2012; Shostak 1995) suggest algorithms capable of discovering radio signals that contain information, a good example being autocorrelation detection (Drake 1965; Harp et al. 2010a; Jones 1995). As implemented here, autocorrelation is sensitive to signals having *arbitrary* message content provided some part of the signal repeats with duration $2 \leq P_n \leq M/4$.

It may be true that somehow, the spectrally pure signal (sinusoid) is strongly favored by builders of ET transmitters over any other signal type, possibly because it mimics spectral line sources. FAC is not at all sensitive to such beacons and the traditional SETI narrowband search is the optimal. However, there are some down sides to sinusoidal signals. The first is that a pure sinusoid carries only one bit of information: that a transmitter exists. Furthermore, it has been argued (Messerschmitt 2012) that sinusoidal signals are less resistant to RFI and multipath (time variable scintillation-induced) fading than say, a spread spectrum signal that is repetitively modulated. In this case a narrowband filter will be relatively insensitive to the signal, yet FAC will be. One last argument against sinusoidal beacons is that most Earth-based radio communications are rapidly converging on spread spectrum signals.

### 4.3  Sensitivity and Choice of Observation Targets

One argument favoring MFB detectors over any other is that they achieve the theoretical limit for sensitivity to a specific waveform. For a fixed transmission power, waveforms associated with a matched filter can be detected from the greatest distance, $R_{max}$. The spherical volume centered on Earth including all transmitter positions detectable at some threshold is largest for the largest values of $R_{max}$. In a blind search, this puts FAC at a great disadvantage compared to PS, since transmitters must be closer to be detected by FAC.

This is why approximately half of our survey was directed toward known continuum sources emitting a high flux. In such cases, we already know that there is enough power arriving from the source to trigger a FAC detector, if that power were modulated. By selecting bright targets, the distance between the source and Earth becomes irrelevant. This leads to the conclusion that FAC detection has the greatest value when applied to bright sources, but has substantially less value than PS for a blind search or survey of the sky.

## 4.4 An Improved FAC Search Design for the Future

We have shown that FAC detection probes an immense region of signal discovery space not observed in conventional SETI campaigns. In designing the next generation search we consider two target types: 1) Search for repetition in radiation from known strong sources, and 2) Search for repetitive signals from directions where no strong flux is expected.

Targets having known flux should be chosen such that they carry sufficient flux to trigger FAC detection. The minimum detectable flux in this paper is ~1 Jy which is well above the threshold for FAC at the ATA. The Very Large Array archives contain measurements on many sources above -20 degrees declination. Using such archive data may allow an expansion of this survey for repetitive signals, but we have not pursued that here.

We speculate that a class of sources might exist that do not show substantial flux over very wide bandwidths (~100 MHz), but do show flux over modest bandwidths (Hz to 10's of MHz). An examination of Table 1 shows that such sources would be more easily discovered using a total power imaging survey than with direct application of FAC processing. This is because the sensitivity to raw power is greater than the sensitivity to FAC.

From these considerations, we propose the next generation SETI detection system should include 3 components. 1) A total power imaging survey covering all frequencies between 1-10 GHz. 2) A conventional narrowband SETI detector. 3) An autocorrelation detector running in parallel and even using the same data streams as for 1) and 2).

## 4.5 Computation of FAC from Correlator Output

FAC can be applied to any observed power spectral data. And besides single-dish or phased array beam data, we believe we are the first to point out that radio interferometer data for imaging can be an excellent source of data for archival FAC searches. Large archives of untested data are available from many telescopes, including the VLA, Westerbork, SRT, GMRT, MeerKat, and ASKAP at centimeter wavelengths; LOFAR and other low frequency arrays; and at high frequencies with ALMA.

Most interferometers designed for radio imaging employ correlators that also produce spectral data. Visibilities are integrated over time periods typically from seconds to minutes, and then stored for subsequent image generation. Such datasets are well suited for FAC because a delay spectrum can be computed from each of hundreds to thousands of correlation spectra. RFI rejection can be implemented by comparing delay spectra across all antenna pairs, since for a point source in the field of view, a repetitive ET signal should have the same FAC in every correlation, whereas RFI coming from outside the field of view will not be constant across correlations.

We demonstrate this approach using archival data from a joint experiment between the ATA and the Arecibo Observatory Figure 5. The Arecibo planetary radar was pointed toward the center of the Moon, and transmissions using binary phase-shift keying (BPSK) were sent with a transmitter bandwidth of 27 MHz. The chip (symbol) duration was 0.5 μs and the same string of 63 symbols was sent over and over. The signal had no amplitude modulation (only phase) with a repetition rate of 31.5 μs.

The ATA was pointed toward the Moon with a field of view much larger than the Moon's diameter. Arecibo transmitted at 2380 MHz with 250 kW of power. Neglecting specular reflection we assume that this power was 100%[10] diffusely scattered into a hemisphere, and taking into account the distance between Earth and Moon, the proportion arriving in the ATA was less than $10^{-7}$ of the transmitted signal. The ATA filling factor is less than 1%, so the signal entering our array aperture was about $2.5 \times 10^{-7}$ W; it was so strong that it was difficult to attenuate the signal enough to prevent overdriving the receivers.

Because the radar is circularly polarized, choosing different polarizations on antenna pairs (XY or YX, cross-polarization) captures the reflected polarized signal (shown) and isolating the radar signal from background radiation. XX, YY spectra (not shown for brevity) showed very similar features. Visibility spectra from the Moon-bounce observation were Fourier transformed to the delay domain and then summed coherently. Clear peaks are seen in Figure 5, with delays corresponding to the 31.5 μs repeat time of the transmitted BPSK signal.

---

[10] The true reflectivity is about 10% of this.

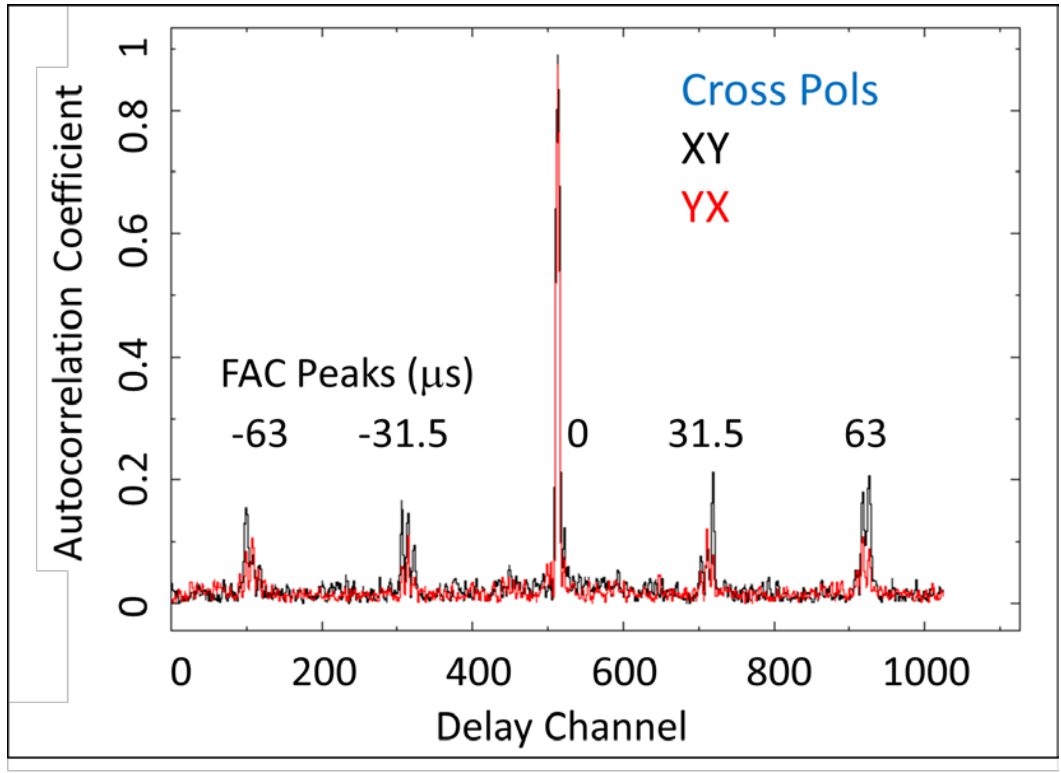

**Figure 4: Demonstration of the FAC technique using an imaging correlator as detector. Since most modern interferometer telescopes use correlators, this method has wide application.**

Compared to the $2 \times 10^6$ delay values examined in this survey, FAC observations with a correlator are limited in delay. The ATA correlator has a fixed number of 1024 spectral bins independent of the spectral bandwidth. Thus, no more than 512 delay values may be observed. Since correlators generally integrate over periods measured in seconds (possibly ms periods in the next generation), we are limited to a maximum delay equal to half the inverse bin width. At the ATA, the maximum delay values for 104 MHz to 6.5 MHz settings are 4.8 μs to 77 μs, respectively. Hence FAC signal searches are optimized when using the maximum number of channels in the correlator, and different delay ranges can be covered with different correlator observation bandwidths.

### 4.6 Beyond Detection of FAC Signals

The relationship between artificial repetition and real sky structure in correlator images deserves further comment. When a point source is within the field of view (FOV), it generally appears as regular oscillations or fringes within the visibility bandwidth for every antenna pair.[11] The fringe period versus frequency depends on the orientation and distance between the antenna pair (baseline), and it is stationary on all baselines when the interferometer is phased up on the source. We call these *structure fringes* because they originate from source structure in the field of view.

---

[11] For discussion, assume the source bandwidth is greater than observation bandwidth.

Repetitive sources introduce what we call *delay fringes* into the visibility spectra. Unless mitigated, delay fringes generate spurious structure in the radio image. Unlike structure fringes, delay fringes have the same oscillation period versus frequency in every visibility, independent of the antenna positions. The latter property allows us to distinguish real spatial structure from signal repetition.

Structure fringes show peaks in the delay domain, one for each point source in the image. Sources at phase center peak at zero delay. Sources far from phase center exhibit peaks with a range of positive or negative delays on different baselines.

In comparison, if a repetitive signal (from anywhere) is present, then FAC peaks will appear at the same delay values for all baselines. Because delay fringes transform to a range of differently scaled structure in images, artificial repetitive signals may generate telescope-specific and time-dependent "halos" around the phase center.

Summing delay fringes over baselines reinforces repetition delay peaks relative to structural delay peaks, at once providing a robust detection scheme for identifying artificial repetitive signals. If these signals are unwanted, their interference peaks may be zeroed out in delay spectra after which a Fourier transform back to the frequency domain can result in improved imaging.

# 5 Discussion

## 5.1 Archetypal Signal Motivating This Work

Almost all modes of human communication are fundamentally symbolic, with symbols repeating in a complex message. This campaign attempts to find signals containing at least some repetitive periodic structure. An archetypal signal here is the wireless Ethernet protocol. Ethernet uses a binary alphabet of symbols to represent arbitrary information. At the hardware level, each symbol is represented by a particular electromagnetic waveform in the radio frequency band. Such signals are transmitted and in principle can escape to great distances from the Earth.

A distant astronomer with a radio telescope and sufficient sensitivity can detect the transmitted data stream in a total power measurement. The astronomer sees broadband power arriving from the direction of Earth. Since many natural sources generate broadband power, the artificial nature of the signal might be overlooked. Alternatively, if she employs a FAC detector then the signal's artificial nature is immediately evident. The FAC detectors are well suited to identify a host of protocols that transmit information using a finite alphabet of symbols.[12]

---

[12] We note that it would be difficult but not impossible to design a symbolic protocol that evades FAC detection.

In this prototype survey we have learned much about how to effectively mount an expanded and more sensitive SETI campaign for detection of repetitive signals.

# 6 Conclusions

This paper reports a novel survey of galactic and extra-galactic targets (quasars, supernova remnants, masers, pulsars, stars with known planets, O-type stars, the Lagrange L4 point, etc.) searching for artificially engineered signals bearing repetitive structure. Signal detection was performed using ~10 minute observations with 7 MHz effective bandwidth captured to disk and post-processed using autocorrelation of the complex-valued voltage signal.

The scientific outcomes of this survey include:
1. Of the known continuum sources with flux > 1 Jy at 1420 MHz, no more than 1 in 20 sources are being modulated.
2. The same is true for stars known to have exoplanets.

When a new probe for extraterrestrial signals is introduced, even a modest survey can add greatly to the human reservoir of knowledge. This paper reports the first search for repetitive extraterrestrial signals, some of which might be hiding "in plain sight" but not detected in ordinary analyses of radio telescope data. This principle is demonstrated by the discovery of heretofore unknown and strong repetitive RFI at the ATA.

We were able to eliminate all observed repetitive signals as candidates for an extraterrestrial signal using direction of origin methods alone and without prior knowledge of human transmitters' frequencies or physical locations. This is generally true for all the SETI performed at the SETI Institute, with the exception of a relatively small number of signals that disappear before their direction of origin can be determined. There were no such cases in the present work.

Finally, we lay the groundwork for a proposed campaign to look for repetitive signals using archived correlator data, especially on known sources with continuum flux at several times the noise floor of the telescope. Such a campaign requires no new observations and can explore all of the sky over a modest range of repetition delays.

## Acknowledgements

The authors gratefully acknowledge Franklin P. Antonio for support of this research. We also thank an anonymous referee for astute comments and helpful discussion which greatly improved the paper during review.

# 7 Appendix

## 7.1 Details of Digital Computation
The actual numerical calculations of PS, FAC and IAC are rather more complex than the notional equations (2), (3) and (4). For all statistics, the initial Fourier transform is performed on a block of $M = 2^{22}$ samples corresponding to one-half second in time. Prior to FFT the block is shifted to zero mean to eliminate the uninteresting zero frequency feature. The block is then multiplied by a Hanning window. This is a raised cosine function that goes to zero at the edges of the sampled time, and removes many artifacts present in a simple periodogram. After the first block, the latter $M/2$ samples from the first block is retained and concatenated with $M/2$ new samples from the dataset. In conjunction with the Hanning window, this overlapping preserves the total power of the sampled data after FFT.

For PS and FAC, we use the sampled electric field values as the starting point for the FFT. After the first FFT, the transformed values are squared to compute the power spectrum (PS). For FAC, the averaged power spectrum is padded with $M/2$ zeros on each end to form an array of length $2M$. This padding prevents mixing of positive and negative delays. The result is passed through an inverse FFT (Weiner-Kinchin theorem) to find the complex-valued autocorrelation statistic. When the signal is stationary, the complex-valued statistic can be averaged over long times. However, this is not always the case (see discussion of Figure 3A and B). To minimize cancellation to the extent possible, we average the absolute value of the FAC statistic over one-half second intervals, hence it is real and positive definite at all points with units of power.

As computed, correlation statistics are biased with a smoothly-varying envelope, peaking at zero delay and going to zero at maximum delay. We correct (normalize) for this bias by dividing each FAC coefficient by the coefficient for a signal containing pure Gaussian noise over a very long time (>>10 minutes) and subjected to the same analysis. Although this normalization introduces mild heteroscedasticity, we focus our attention on only the lower 50% of delay values to minimize its effects.

The raw autocorrelation values in the plots of Figure 3 have $2^{21}$ individual points, much more than can be adequately displayed in a printed plot. To bring the number of points down to a manageable group, the original AC data are binned with a max-pooling method (each point represents the largest of 1024 points in the raw AC data). To give a better impression of the actual point density we blow up a region around the first peak in Figure 3B which includes all the computed AC values between 6000 and 7000 microseconds.

The convolution algorithm (Wiener–Khinchin theorem) for FAC is important in this application as it costs only 2x more computation than for the PS (Harp et al. 2010a). A more sensitive FAC-based algorithm, symbol-wise autocorrelation (Morrison 2012, 2017) might be pursued in future studies but will require >400,000 times more computation time than for the campaign presented here.

## 7.2 Appendix: Observation Table

List of all observations collected for this project. The raw data for these observations are available online (Harp et al. 2010b). Sources with "off" in their name track the path of the stated object but following the real source with a delay of ~ 1 hr.

Table 3:

List of Observations for This Study

| Source | Coordinate | Type | Freq (MHz) | Flux (Jy) @obs freq | Date |
|---|---|---|---|---|---|
| 0136+478 | 01 36 59 +47 51 29 | quasar | 2008 | 1.7 | 2010-12-24 |
| 0136+478 | 01 36 59 +47 51 29 | quasar | 4462 | 1.9 | 2010-12-17 |
| 0136+478 | 01 36 59 +47 51 29 | quasar | 6670 | 1.9 | 2010-12-12 |
| 0228+673 | 02 28 50 +67 21 03 | quasar | 2008 | 1.2 | 2011-01-07 |
| 0228+673 | 02 28 50 +67 21 03 | quasar | 6670 | 2.0 | 2010-12-12 |
| 0744-064 | 07 44 22 -06 29 36 | quasar | 2840 | 4.7 | 2010-07-02 |
| 0834+555 | 08 34 55 +55 34 21 | quasar | 2840 | 6.8 | 2010-07-02 |
| 0834+555 | 08 34 55 +55 34 21 | quasar | 2840 | 6.8 | 2010-12-24 |
| 0834+555 | 08 34 55 +55 34 21 | quasar | 3100 | 6.6 | 2010-07-22 |
| 0834+555 | 08 34 55 +55 34 21 | quasar | 6670 | 4.4 | 2010-12-17 |
| 1347+122 | 13 47 33 +12 17 24 | quasar | 2008 | 4.6 | 2010-12-24 |
| 1347+122 | 13 47 33 +12 17 24 | quasar | 4462 | 3.0 | 2010-12-24 |
| 1347+122 | 13 47 33 +12 17 24 | quasar | 6670 | 2.7 | 2010-12-12 |
| 1733-130 | 17 33 03 -13 04 50 | quasar | 2008 | 5.1 | 2011-02-04 |
| 1733-130 | 17 33 03 -13 04 50 | quasar | 4462 | 4.8 | 2010-12-17 |
| 1733-130 | 17 33 03 -13 04 50 | quasar | 6670 | 10.1 | 2010-12-12 |
| 2038+513 | 20 38 37 +51 19 13 | quasar | 2008 | 4.9 | 2011-02-04 |
| 2038+513 | 20 38 37 +51 19 13 | quasar | 6670 | 3.4 | 2010-12-17 |
| 2206-185 | 22 06 10 -18 35 39 | quasar | 2008 | 5.8 | 2011-02-04 |
| 2206-185 | 22 06 10 -18 35 39 | quasar | 4462 | 3.9 | 2010-12-17 |
| 3c119 | 04 32 37 +41 38 28 | quasar | 2008 | 7.1 | 2011-02-04 |
| 3c119 | 04 32 37 +41 38 28 | quasar | 2840 | 5.4 | 2010-07-02 |
| 3c119 | 04 32 37 +41 38 28 | quasar | 4462 | 3.9 | 2010-06-25 |
| 3c123 | 04 37 04 +29 40 14 | quasar | 2008 | 39.7 | 2011-02-04 |
| 3c123 | 04 37 04 +29 40 14 | quasar | 2840 | 31.7 | 2010-07-02 |
| 3c123 | 04 37 04 +29 40 14 | quasar | 4465 | 23.6 | 2010-06-25 |
| 3c138 | 05 21 10 +16 38 22 | quasar | 2008 | 7.1 | 2011-02-04 |
| 3c138 | 05 21 10 +16 38 22 | quasar | 2840 | 5.4 | 2010-07-02 |
| 3c138 | 05 21 10 +16 38 22 | quasar | 4462 | 4.0 | 2010-06-25 |
| 3c147 | 05 42 36 +49 51 07 | quasar | 2008 | 18.1 | 2011-02-04 |
| 3c147 | 05 42 36 +49 51 07 | quasar | 2840 | 12.5 | 2010-07-02 |

| Name | Coordinates | Type | Freq | Flux | Date |
|---|---|---|---|---|---|
| 3c147 | 05 42 36 +49 51 07 | quasar | 4462 | 8.6 | 2010-06-25 |
| 3c274 | 12 30 49 +12 23 29 | quasar | 2008 | 3.0 | 2010-12-24 |
| 3c274 | 12 30 49 +12 23 29 | quasar | 2840 | 3.0 | 2010-07-02 |
| 3c274 | 12 30 49 +12 23 29 | quasar | 4462 | 3.0 | 2010-12-24 |
| 3c274 | 12 30 49 +12 23 29 | quasar | 6670 | 3.0 | 2010-12-17 |
| 3c279 | 12 56 11 -05 47 22 | quasar | 6670 | 13.6 | 2010-12-17 |
| 3c286 | 13 31 08 +30 30 33 | quasar | 1420 | 14.9 | 2010-10-15 |
| 3c286 | 13 31 08 +30 30 33 | quasar | 2008 | 12.9 | 2011-01-07 |
| 3c286 | 13 31 08 +30 30 33 | quasar | 2008 | 12.9 | 2010-12-24 |
| 3c286 | 13 31 08 +30 30 33 | quasar | 4020 | 8.4 | 2010-10-08 |
| 3c286 | 13 31 08 +30 30 33 | quasar | 6670 | 6.3 | 2010-12-12 |
| 3c295 | 14 11 21 +52 12 09 | quasar | 2008 | 17.0 | 2011-01-07 |
| 3c295 | 14 11 21 +52 12 09 | quasar | 6670 | 3.0 | 2010-12-12 |
| 3c345 | 16 42 59 +39 48 37 | quasar | 2008 | 7.9 | 2011-01-07 |
| 3c345 | 16 42 59 +39 48 37 | quasar | 2008 | 7.9 | 2011-02-04 |
| 3c345 | 16 42 59 +39 48 37 | quasar | 6670 | 5.5 | 2010-12-12 |
| 3c380 | 18 29 32 +48 44 46 | quasar | 2008 | 10.6 | 2011-02-04 |
| 3c380 | 18 29 32 +48 44 46 | quasar | 6670 | 4.0 | 2010-12-17 |
| 3c395 | 19 02 56 +31 59 42 | quasar | 6670 | 1.2 | 2010-12-17 |
| 3c48 | 01 37 41 +33 09 35 | quasar | 1422 | 16.5 | 2010-08-13 |
| 3c48 | 01 37 41 +33 09 35 | quasar | 2008 | 13.0 | 2010-12-24 |
| 3c48 | 01 37 41 +33 09 35 | quasar | 2840 | 8.9 | 2010-07-02 |
| 3c48 | 01 37 41 +33 09 35 | quasar | 4462 | 5.9 | 2010-12-17 |
| 3c48 | 01 37 41 +33 09 35 | quasar | 6670 | 4.2 | 2010-12-24 |
| 3c84 | 03 19 48 +41 30 42 | quasar | 2840 | 23.6 | 2010-07-02 |
| 3c84 | 03 19 48 +41 30 42 | quasar | 4462 | 23.4 | 2010-06-25 |
| 3c84 | 03 19 48 +41 30 42 | quasar | 4462 | 23.4 | 2010-09-10 |
| 3c84 | 03 19 48 +41 30 42 | quasar | 6670 | 22.1 | 2010-12-17 |
| **Blazars** | | | | | |
| BLLacterus | 22 02 43 +42 16 40 | blazar | 1420 | 6.1 | 2010-10-15 |
| BLLacterus | 22 02 43 +42 16 40 | blazar | 2008 | 5.9 | 2010-10-15 |
| BLLacterus | 22 02 43 +42 16 40 | blazar | 6670 | 4.2 | 2010-05-21 |
| BLLacterus | 22 02 43 +42 16 40 | blazar | 8200 | 4.0 | 2010-08-13 |
| S5 0716+714 | 07 21 53 +71 20 36 | blazar | 1414 | ~1.5 varies | 2010-10-15 |
| S5 0716+714 | 07 21 53 +71 20 36 | blazar | 1420 | ~1.5 varies | 2010-10-15 |
| S5 0716+714 | 07 21 53 +71 20 36 | blazar | 1420 | ~1.5 varies | 2010-05-21 |
| S5 0716+714 | 07 21 53 +71 20 36 | blazar | 1422 | ~1.5 varies | 2010-08-13 |
| S5 0716+714 | 07 21 53 +71 20 36 | blazar | 1422 | ~1.5 varies | 2010-08-13 |
| S5 0716+714 | 07 21 53 +71 20 36 | blazar | 1422 | ~1.5 varies | 2010-10-15 |
| S5 0716+714 | 07 21 53 +71 20 36 | blazar | 1426 | ~1.5 varies | 2010-08-13 |
| S5 0716+714 | 07 21 53 +71 20 36 | blazar | 1427 | ~1.5 varies | 2010-10-15 |
| S5 0716+714 | 07 21 53 +71 20 36 | blazar | 1432 | ~1.5 varies | 2010-08-13 |
| S5 0716+714 | 07 21 53 +71 20 36 | blazar | 1435 | ~1.5 varies | 2010-05-21 |
| S5 0716+714 | 07 21 53 +71 20 36 | blazar | 1459 | ~1.5 varies | 2010-10-22 |
| S5 0716+714 | 07 21 53 +71 20 36 | blazar | 3086 | ~1.5 varies | 2010-12-17 |

| Name | Coordinates | Type | Freq (MHz) | Flux (Jy) | Date |
|---|---|---|---|---|---|
| S5 0716+714 | 07 21 53 +71 20 36 | blazar | 6670 | ~1.5 varies | 2010-10-22 |
| S5 0716+714 | 07 21 53 +71 20 36 | blazar | 8200 | ~1.5 varies | 2010-05-07 |
| 0954+658 | 09 58 47 +65 33 55 | blazar | 1432 | ~1.3 varies | 2010-10-15 |
| 0954+658 | 09 58 47 +65 33 55 | blazar | 8200 | ~1.3 varies | 2010-10-22 |
| **Supernova Remnants and Masers** | | | | | |
| taua | 05 34 32 +22 00 58 | SNR | 2008 | 1110.0 | 2010-08-13 |
| taua | 05 34 32 +22 00 58 | SNR | 1420 | 1110.0 | 2010-05-07 |
| taua | 05 34 32 +22 00 58 | SNR | 2840 | 1110.0 | 2010-10-15 |
| casa | 23 23 27 +58 48 28 | SNR | 2008 | 1547.3 | 2010-08-13 |
| casa | 23 23 27 +58 48 28 | SNR | 4462 | 885.9 | 2010-10-15 |
| casa | 23 23 27 +58 48 28 | SNR | 6670 | 701.1 | 2010-08-13 |
| w51 | 19 23 44 +14 30 33 | maser | 6670 | 850 peak | 2010-05-07 |
| w3oh | 02 27 04 +61 52 25 | maser | 6670 | 3741 peak | 2010-03-26 |
| w3oh | 02 27 04 +61 52 25 | maser | 6670 | 3741 peak | 2010-04-02 |
| w3oh | 02 27 04 +61 52 25 | maser | 6670 | 3741 peak | 2010-05-07 |
| **Pulsars** | | | | @1400 MHz | |
| crab | 05 34 32 +22 00 52 | pulsar | 2008 | 12.6 | 2011-02-04 |
| crab | 05 34 32 +22 00 52 | pulsar | 1420 | 12.6 | 2010-03-26 |
| crab | 05 34 32 +22 00 52 | pulsar | 1420 | 12.6 | 2010-10-15 |
| crab | 05 34 32 +22 00 52 | pulsar | 2600 | 12.6 | 2010-03-09 |
| crab | 05 34 32 +22 00 52 | pulsar | 4462 | 12.6 | 2010-05-07 |
| crab | 05 34 32 +22 00 52 | pulsar | 1420 | 12.6 | 2010-05-07 |
| crab | 05 34 32 +22 00 52 | pulsar | 4462 | 12.6 | 2010-06-25 |
| psrb0329+54 | 03 32 59 +54 34 44 | pulsar | 1420 | 1.8 | 2010-05-07 |
| psrb0329+54 | 03 32 59 +54 34 44 | pulsar | 1420 | 1.8 | 2010-05-21 |
| psrb0329+54 | 03 32 59 +54 34 44 | pulsar | 611 | 1.8 | 2011-03-04 |
| psrb0450+55 | 04 54 08 +55 43 42 | pulsar | 4462 | 0.3 | 2010-06-25 |
| psrb0809+74 | 08 14 59 +74 29 06 | pulsar | 4462 | 0.0 | 2010-06-25 |
| psrb0809+74 | 08 14 59 +74 29 06 | pulsar | 1420 | 0.0 | 2010-05-14 |
| psrb0823+26 | 08 26 51 +26 37 24 | pulsar | 4462 | 0.1 | 2010-06-18 |
| psrb0823+26 | 08 26 51 +26 37 24 | pulsar | 4462 | 0.1 | 2010-06-25 |
| psrb0950+08 | 09 53 09 +07 55 36 | pulsar | 4462 | 3.2 | 2010-06-18 |
| psrb0950+08 | 09 53 09 +07 55 36 | pulsar | 4462 | 3.2 | 2010-06-25 |
| psrb1133+16 | 11 36 03 +15 51 04 | pulsar | 4462 | 0.9 | 2010-06-25 |
| psrb1237+25 | 12 39 40 +24 53 49 | pulsar | 4462 | 0.4 | 2010-06-25 |
| psrb1937+21 | 19 39 39 +21 34 59 | pulsar | 1420 | 0.3 | 2010-11-05 |
| **Stars - Including Kepler Exoplanets, O-stars** | | | | | |
| koi001 | 13 12 44 -31 52 24 | exoplanet | 4462 | <1 | 2010-09-10 |
| koi044 | 22 53 13 -14 15 13 | exoplanet | 4462 | <1 | 2010-03-26 |
| koi051 | 09 34 50 -12 07 46 | exoplanet | 4462 | <1 | 2010-09-10 |
| koi054 | 10 58 28 -10 46 13 | exoplanet | 4462 | <1 | 2010-09-10 |
| koi060 | 03 32 55 -09 27 29 | exoplanet | 1420 | <1 | 2010-03-19 |
| koi062 | 12 44 20 -08 40 17 | exoplanet | 4462 | <1 | 2010-09-10 |
| koi071 | 11 35 52 -04 45 21 | exoplanet | 4462 | <1 | 2010-09-10 |
| koi072 | 22 09 40 -04 38 27 | exoplanet | 4462 | <1 | 2010-03-26 |
| koi073 | 09 56 06 -03 48 30 | exoplanet | 4462 | <1 | 2010-09-10 |
| koi076 | 12 19 13 -03 19 11 | exoplanet | 4462 | <1 | 2010-09-10 |
| koi080 | 13 12 43 -02 15 54 | exoplanet | 4462 | <1 | 2010-09-10 |
| koi081 | 10 42 48 -02 11 01 | exoplanet | 4462 | <1 | 2010-09-10 |
| koi084 | 11 24 17 -01 31 44 | exoplanet | 4462 | <1 | 2010-09-10 |
| koi096 | 11 45 42 +02 49 17 | exoplanet | 4462 | <1 | 2010-09-10 |
| koi097 | 11 26 46 +03 00 22 | exoplanet | 4462 | <1 | 2010-09-10 |
| koi112 | 12 13 29 +10 02 29 | exoplanet | 4462 | <1 | 2010-09-10 |
| koi117 | 10 18 21 +12 37 15 | exoplanet | 4462 | <1 | 2010-09-10 |
| koi118 | 13 28 26 +13 47 12 | exoplanet | 4462 | <1 | 2010-09-10 |
| koi119 | 11 46 24 +14 07 26 | exoplanet | 4462 | <1 | 2010-09-10 |
| koi125 | 13 12 19 +17 31 01 | exoplanet | 4462 | <1 | 2010-09-10 |
| koi126 | 10 10 07 +18 11 12 | exoplanet | 4462 | <1 | 2010-09-10 |
| koi127 | 23 18 47 +18 38 45 | exoplanet | 3100 | <1 | 2010-07-22 |
| koi130 | 04 42 56 +18 57 29 | exoplanet | 1420 | <1 | 2010-03-19 |
| koi133 | 09 23 47 +20 21 52 | exoplanet | 3100 | <1 | 2010-07-22 |
| koi134 | 00 44 41 +20 26 56 | exoplanet | 4462 | <1 | 2010-03-26 |
| koi144 | 00 39 21 +21 15 01 | exoplanet | 1420 | <1 | 2010-03-19 |
| koi150 | 02 04 34 +25 24 51 | exoplanet | 4462 | <1 | 2010-03-26 |
| koi155 | 11 42 11 +26 42 23 | exoplanet | 4462 | <1 | 2010-09-10 |
| koi160 | 08 52 37 +28 20 02 | exoplanet | 1420 | <1 | 2010-03-19 |
| koi172 | 00 20 40 +31 59 24 | exoplanet | 4462 | <1 | 2010-03-26 |
| koi191 | 22 57 47 +38 40 30 | exoplanet | 3100 | <1 | 2010-07-22 |
| koi195 | 10 59 29 +40 25 46 | exoplanet | 4462 | <1 | 2010-09-10 |
| koi201 | 10 22 10 +41 13 46 | exoplanet | 4462 | <1 | 2010-09-10 |
| koi201 | 10 22 10 +41 13 46 | exoplanet | 3100 | <1 | 2010-07-22 |
| koi202 | 01 36 48 +41 24 38 | exoplanet | 1420 | <1 | 2010-03-19 |
| koi208 | 06 04 29 +44 15 37 | exoplanet | 3100 | <1 | 2010-07-22 |
| koi211 | 19 28 59 +47 58 10 | exoplanet | 1420 | <1 | 2010-09-24 |
| koi211 | 19 28 59 +47 58 10 | exoplanet | 4462 | <1 | 2010-10-01 |
| koi216 | 07 48 07 +50 13 33 | exoplanet | 3100 | <1 | 2010-07-22 |
| koi219 | 14 56 55 +53 22 56 | exoplanet | 1420 | <1 | 2010-09-24 |
| koi219 | 14 56 55 +53 22 56 | exoplanet | 4462 | <1 | 2010-10-01 |
| koi220 | 13 34 02 +53 43 42 | exoplanet | 4462 | <1 | 2010-09-10 |
| koi221 | 15 35 16 +53 55 20 | exoplanet | 1420 | <1 | 2010-09-24 |
| koi221 | 15 35 16 +53 55 20 | exoplanet | 4462 | <1 | 2010-10-01 |
| koi222 | 18 10 32 +54 17 12 | exoplanet | 4462 | <1 | 2010-10-01 |
| koi225 | 07 21 33 +58 16 05 | exoplanet | 4462 | <1 | 2010-10-01 |
| koi226 | 15 24 55 +58 57 57 | exoplanet | 4462 | <1 | 2010-10-01 |
| koi227 | 08 18 22 +61 27 38 | exoplanet | 4462 | <1 | 2010-10-01 |
| koi228 | 08 40 13 +64 19 41 | exoplanet | 4462 | <1 | 2010-10-01 |
| koi231 | 12 05 15 +76 54 20 | exoplanet | 4462 | <1 | 2010-09-10 |
| koi233 | 05 22 33 +79 13 52 | exoplanet | 3100 | <1 | 2010-07-22 |
| koi236 | 13 00 03 +12 00 07 | exoplanet | 4462 | <1 | 2010-09-10 |

| Name | Coordinates | Type | Freq (MHz) | SEFD | Date |
|---|---|---|---|---|---|
| Gliese581 | 15 19 27 -07 43 20 | exoplanet | 1420 | <1 | 2010-10-01 |
| Gliese581 | 15 19 27 -07 43 20 | exoplanet | 4462 | <1 | 2011-01-28 |
| Gliese581 | 15 19 27 -07 43 20 | exoplanet | 4462 | <1 | 2010-10-01 |
| koi04 | 19 02 28 +50 08 09 | exoplanet | 1418 | <1 | 2010-05-14 |
| koi04 | 19 02 28 +50 08 09 | exoplanet | 1420 | <1 | 2010-01-22 |
| koi04 | 19 02 28 +50 08 09 | exoplanet | 1420 | <1 | 2010-03-19 |
| koi04 | 19 02 28 +50 08 09 | exoplanet | 1420 | <1 | 2010-03-19 |
| koi04 | 19 02 28 +50 08 09 | exoplanet | 1420 | <1 | 2010-04-02 |
| koi04 | 19 02 28 +50 08 09 | exoplanet | 1420 | <1 | 2010-09-24 |
| koi04 | 19 02 28 +50 08 09 | exoplanet | 1420 | <1 | 2010-11-05 |
| koi04 | 19 02 28 +50 08 09 | exoplanet | 1420 | <1 | 2010-05-14 |
| koi04 | 19 02 28 +50 08 09 | exoplanet | 1692 | <1 | 2010-09-24 |
| koi139.01 | 19 26 37 +44 41 18 | exoplanet | 1690 | <1 | 2011-03-31 |
| koi174.01 | 19 47 18 +48 06 27 | exoplanet | 1690 | <1 | 2011-03-31 |
| koi268.01 | 19 02 55 +38 30 25 | exoplanet | 1690 | <1 | 2011-03-31 |
| koi51.01 | 19 43 40 +41 19 57 | exoplanet | 1690 | <1 | 2011-03-31 |
| koi70.03 | 19 10 48 +42 20 19 | exoplanet | 1690 | <1 | 2011-03-31 |
| 55Cancri | 08 52 36 +28 19 51 | OZMA star | 1420 | <1 | 2010-11-05 |
| zetaOph | 16 37 10 -10 34 02 | O-star | 1420 | <1 | 2010-11-05 |
| zetaOph | 16 37 10 -10 34 02 | O-star | 2008 | <1 | 2010-12-24 |
| zetaOph | 16 37 10 -10 34 02 | O-star | 2840 | <1 | 2010-12-24 |
| zetaOph | 16 37 10 -10 34 02 | O-star | 6670 | <1 | 2010-12-17 |
| HD172175 | 18 39 04 -07 51 35 | O-star | 1420 | <1 | 2010-11-05 |
| HD166734 | 18 12 25 -10 43 53 | O-star | 1420 | <1 | 2010-11-05 |
| HD093521 | 10 48 24 +37 34 13 | O-star | 1420 | <1 | 2010-11-05 |
| HD060848 | 07 37 06 +16 54 15 | O-star | 1420 | <1 | 2010-11-05 |
| BD114586 | 18 18 03 -11 17 39 | O-star | 1420 | <1 | 2010-11-05 |
| **Special Pointings** | | | | | |
| Sun | N/A | Sun | 1414 | ~1000000 | 2010-10-15 |
| Sun | N/A | Sun | 1420 | ~1000000 | 2010-10-15 |
| Sun | N/A | Sun | 1426 | ~1000000 | 2010-10-15 |
| Moon | N/A | Moon | 1420 | ~10000 | 2010-10-08 |
| Moon | N/A | Moon | 1420 | ~10000 | 2010-11-05 |
| northpole | 00 00 00 +90 00 00 | North Pole | 1543 | <1 | 2010-10-22 |
| northpole | 00 00 00 +90 00 00 | North Pole | 3086 | <1 | 2010-10-22 |
| Lagrange-4 | Sun-Earth L4 | Lagrange point | 1420 | <1 | 2010-10-08 |
| Lagrange-4 | Sun-Earth L4 | Lagrange point | 2008 | <1 | 2010-10-08 |
| Lagrange-4 | Sun-Earth L4 | Lagrange point | 3991 | <1 | 2010-10-08 |
| galanticenter | 05 45 37 +28 56 10 | Galactic Anticenter | 1420 | <1 | 2010-05-07 |
| galanticenter | 05 45 37 +28 56 10 | Galactic Anticenter | 1422 | <1 | 2010-05-07 |

| Name | Coordinates | Type | Freq | ? | Date |
|---|---|---|---|---|---|
| galanticenter | 05 45 37 +28 56 10 | Galactic Anticenter | 3991 | <1 | 2010-05-07 |
| galanticenter | 05 45 37 +28 56 10 | Galactic Anticenter | 3991 | <1 | 2010-10-08 |
| galanticenter | 05 45 37 +28 56 10 | Galactic Anticenter | 4462 | <1 | 2010-05-07 |
| galanticenter | 05 45 37 +28 56 10 | Galactic Anticenter | 2008 | <1 | 2010-05-07 |
| galanticenter | 05 45 37 +28 56 10 | Galactic Anticenter | 3991 | <1 | 2010-10-08 |
| **Confirmation Observations (intentionally away from some particular source)** | | | | | |
| blank06 | 06 00 00 +40 00 00 | off beam | 1422 | | ? |
| blank06 | 06 00 00 +40 00 00 | off beam | 1427 | | ? |
| blank18 | 18 00 00 +70 00 00 | off beam | 6670 | | ? |
| blank18 | 18 00 00 +70 00 00 | off beam | 2008 | | ? |
| blank | 09 18 06 -11 54 17 | off beam | 2840 | | ? |
| off EpsilonEridani | 03 32 56 +09 27 30 | off beam | 1420 | | ? |
| off EtaAreitis | 02 12 48 +21 12 39 | off beam | 1420 | | ? |
| off galanticenter | 05 45 37 +33 56 10 | off beam | 2008 | | ? |
| off galanticenter | 05 45 37 +33 56 10 | off beam | 3991 | | ? |
| off Gliese581 | 15 19 27 +07 43 20 | off beam | 4462 | | ? |
| off Gliese581 | 15 19 27 +07 43 20 | off beam | 1420 | | ? |
| off HD69830 | 08 18 24 +12 37 56 | off beam | 1420 | | ? |
| off TauCeti | 01 44 04 +15 56 15 | off beam | 1420 | | ? |
| **Confirmation Observations (surveying RFI at ATA)** | | | | | |
| AZEL360-18 | fixed az, el | Az sweep | 1410 | | ? |
| AZEL345-18 | fixed az, el | Az sweep | 1410 | | ? |
| AZEL330-18 | fixed az, el | Az sweep | 1410 | | ? |
| AZEL315-18 | fixed az, el | Az sweep | 1410 | | ? |
| AZEL300-18 | fixed az, el | Az sweep | 1410 | | ? |
| AZEL285-18 | fixed az, el | Az sweep | 1410 | | ? |
| AZEL270-18 | fixed az, el | Az sweep | 1410 | | ? |
| AZEL255-18 | fixed az, el | Az sweep | 1410 | | ? |
| AZEL240-18 | fixed az, el | Az sweep | 1410 | | ? |
| AZEL225-18 | fixed az, el | Az sweep | 1410 | | ? |
| AZEL210-18 | fixed az, el | Az sweep | 1410 | | ? |
| AZEL195-18 | fixed az, el | Az sweep | 1410 | | ? |
| AZEL180-18 | fixed az, el | Az sweep | 1410 | | ? |
| AZEL165-18 | fixed az, el | Az sweep | 1410 | | ? |
| AZEL150-18 | fixed az, el | Az sweep | 1410 | | ? |
| AZEL135-18 | fixed az, el | Az sweep | 1410 | | ? |
| AZEL120-18 | fixed az, el | Az sweep | 1410 | | ? |
| AZEL105-18 | fixed az, el | Az sweep | 1410 | | ? |

| | | | | |
|---|---|---|---|---|
| AZEL090-18 | fixed az, el | Az sweep | 1410 | ? |
| AZEL075-18 | fixed az, el | Az sweep | 1410 | ? |
| AZEL060-18 | fixed az, el | Az sweep | 1410 | ? |
| AZEL045-18 | fixed az, el | Az sweep | 1410 | ? |
| AZEL030-18 | fixed az, el | Az sweep | 1410 | ? |
| AZEL015-18 | fixed az, el | Az sweep | 1410 | ? |